# MARTINI-based force fields for predicting gas separation performances of MOF/polymer composites

Cecilia M. S. Alvares,[1] Rocio Semino[2*]

[1] ICGM, Univ. Montpellier, CNRS, ENSCM, Montpellier, France
[2] Sorbonne Université, CNRS, Physico-chimie des Electrolytes et Nanosystèmes Interfaciaux, PHENIX, F-75005 Paris, France
* rocio.semino@sorbonne-universite.fr

**Abstract**

MOF/polymer composites have been widely investigated in the past decade for gas separation applications. However, the impact of MOF nanoparticle morphology and size in gas separation have not yet been systematically studied by computer simulation techniques. In this work, coarse grained simulations are deployed to study gas adsorption in ZIF-8/PVDF at the nanoparticle level. Nanoparticles of different morphologies and sizes are explored, and adsorption of $CO_2$, $N_2$ and $CH_4$ is investigated throughout the extension of the bulk and surface of the nanoparticle as well as of the polymer phase. Results reproduce the expected preference for $CO_2$ over the other two gasses. Nanoparticles of smaller sizes provide better separation performance at ambient conditions, while rhombic dodecahedron nanoparticles perform better than cubic ones. This work presents a perspective on the merits and limitations of modelling gas adsorption in MOF/polymer composites at the nanoparticle level via particle-based coarse graining approaches, and provides a methodological set-up which can be integrated into high-throughput schemes, bringing us closer to reaching time- and length scales that can be directly compared with experimental data.

**I) Introduction**

In the past decade, metal-organic frameworks (MOFs) membrane technologies have been augmented by combining them with polymers, which has improved not only their stability[1] and mechanical properties,[2,3] but also the performance over pure polymer membranes for diverse societally and industrially relevant applications,[4-7] including challenging gas separations, such as pre- and post-combustion $CO_2$ separations ($CO_2/CH_4$ and $CO_2/N_2$ respectively).[8] In addition to the large library of MOFs featuring different pore sizes, topologies, and chemical functions, a large variety of polymers can be employed to build these composite materials. However, tuning these mixed matrix membranes for improving their performance has proven challenging. While there is wide consensus that the structural characteristics of MOF/polymer interfaces play key roles in determining membrane performance,[9] no general rules have yet been established. Indeed, even though many strategies have been applied to improve $CO_2/CH_4$ gas separation merit parameters,[10] it seems that the best technique to tailor the MOF/polymer interface for gas separation applications is case dependent.[11] Historically, focus has been put on choosing a good MOF/polymer combination.[10] "Good" can be defined as having no additional porosity at the interface and/or by a high chemical affinity or compatibility. However, "good" or "poor" compatibilities are not directly correlated with gas separation performances. Good compatibility resulting from a strong MOF-polymer adhesion may lead to pore blocking and a subsequent reduction in the accessible porosity of the MOF.[12,13] Poor compatibility, in turn, may be regarded as interfacial defects, but those defects are not necessarily detrimental for applications. Indeed, Fan and coworkers determined that interfacial voids may help in gas separation applications, by acting as a reservoir of gas molecules directly at the entrance of the MOF pores, for a MOF that separates gasses through a mechanism controlled by a solubility difference.[14] It can thus be concluded from these results that it is not possible to make any prediction about performance for applications from compatibility alone. The fact that defects can contribute favorably to gas separation performances is at the basis of another set of strategies that do not involve getting rid of defects, but introducing and engineering them instead.[15,16]

Computational works have been fundamental to the study of adsorption phenomena in porous materials,[17,18] with methods ranging from DFT calculations[19] to Grand Canonical Monte Carlo (GCMC) or molecular dynamics (MD) simulations,[20-25] spanning from all-atom up to different kinds of coarser models.[26,27] Composite materials based on porous solids have also been the object of many computational studies, which are particularly important given the difficulty in obtaining molecular level resolution details of interfaces through experimental measurements. The pioneering work by Zhang and collaborators has enabled reproducing experimental gas separation performances trends even with a very simple model of the interface.[28] Permeation models combined with atomistic simulations have enabled the Keskin group to perform high-throughput screening of MOF/polymer composites to find the most promising pairs in terms of their merit parameters.[29,30] Semino and coworkers developed a microscopic methodology to model MOF/polymer interfaces at the all-atom resolution that allowed to study compatibility at the molecular level.[31] Alvares and Semino studied the structure of ZIF-8/PVDF composites having a rhombic dodecahedron or a cubic ZIF-8 nanoparticle immersed in the polymer matrix.[32] They found that the polymer penetrates through the entire extension of the ZIF-8 phase in all cases, albeit with lower local density values than those of the bulk polymer, and that gas still is preferentially adsorbed in the nanoparticle domain compared to the polymer phase, in alignment with experimental expectations.

Despite these important steps forward in the field, crucial questions that can only be answered by modelling the system at larger scales still remain, such as what is the impact of nanoparticle size and morphology on gas separation? Deng and coworkers have studied the impact of ZIF morphology in $CO_2$ separation performances from the experimental point of view and found that the lower the ZIF filler dimensionality, the better the separation.[33] Additionally, aligning the MOF filler within the polymer matrix[34,35] is a useful strategy to improve performances, and it could also explain the reason behind the better separation performances afforded by 2D fillers when compared to 3D ones.[36] However, in a study where a series of MOF and polymers were scanned to produce mixed matrix membranes (MMMs), Sabetghadam and coworkers found composites where MOFs were included in the form of nanoparticles yielded higher gas separation performances than composites where MOFs were included in the form of nanorods and microneedles.[37] Furthermore, adsorption properties and selectivities can change significantly with crystal size.[38,39] The $CO_2/N_2$ selectivity was found to increase with nanoparticle size in ZIF-8/PEBAX MMMs[40] and the same tendency was obtained for $CO_2/H_2$ selectivity for ZIF-8/PBI membranes.[41] However, Chi and coworkers found that middle sized particles yielded higher selectivities towards $CO_2$ for ZIF-8/SEBS block copolymer composites.[42] Still on the matter of performance, it was found that smaller ZIF-8 nanoparticles in a ZIF-8/polyamide thin film membrane lead to better performance in reverse osmosis processes.[43] Similar conclusions about smaller-sized MOF particles have been found when using ZIF-90 and ZIF-71 based MOF/polymer membranes for separating $CO_2/CH_4$ and several gas pairs ($H_2/CH_4$, $H_2/N_2$, $H_2,CO_2$), respectively.[44,45] A link between smaller sized nanoparticles and better performance has also been made for different MOF/polyimide composites for ethylene/ethane separation.[46] These apparently contradictory findings regarding particle size and performance prove that the topic is not a simple one, and could use the help of computer simulations.

In this work, we offer insights on the impact of nanoparticle size and morphology on gas separation by computationally studying single gas adsorption within a highly compatible MOF/polymer pair: ZIF-8/PVDF. PVDF stands for polyvinylidene fluoride, a polymer widely used in the membrane community. ZIF-8, a MOF formed by $Zn^{2+}$ and 2-methylimidazolate ligands, is particularly interesting for its high chemical stability and numerous applications.[47] The ZIF-8/PVDF pair has raised interest for many different applications including challenging gas separations.[48-50] In what follows, particle-based coarse graining models are used to model ZIF-8/PVDF composites at the nanoparticle scale. We employ potentials coming from a MARTINI-3[51]/Force Matching (FM)[52] ZIF-8/PVDF model from our previous work,[32] and we add MARTINI 3 potentials to account for the gas interactions. We show that MARTINI 3 potentials can be successfully used to model gas adsorption in binary MOF/gas systems, with results that are comparable with experimental trends. Given the success and simplicity of our proposed methodology, we anticipate that it may be used to perform systematic studies over many MOF/polymer pairs to assess the influence of nanoparticle size and shape in selectivity.

This article is organized as follows: section II presents the methods, including the force field details, section III contains the results and discussions and section IV summarizes the conclusions.

**II) Methodology**

All the simulations involved in this work were made using LAMMPS.[53] Molecular dynamics simulations were carried out using the Nosé Hoover equations of motion as originally developed to either sample the NVT or the NPT ensemble.[54] These are referred to as "NVT equations of motion" and "NPT equations of motion", respectively, throughout the text. Damping constants for the thermostat and (when applicable) for the barostat were always of 100x and 1000x the value of the timestep used in the simulation, respectively.

II.1) CG model development

ZIF-8/PVDF composites under single gas loading were computationally modelled at the mesoscale level (simulation domains ≈20 nm large in x, y, z directions) in the CG resolution in order to get insights on the influence of the filler size and shape in $CO_2$, $CH_4$ and $N_2$ adsorption and structuring at ambient (T,P) conditions. Since all simulation setups involve single gas loading and three different gasses are considered, three ternary systems are studied: ZIF-8/PVDF/$CO_2$, ZIF-8/PVDF/$CH_4$ and ZIF-8/PVDF/$N_2$. Naturally, modelling each of these systems at the CG level requires defining a mapping as well as having a force field. The mapping for the ZIF-8 and PVDF phases is shared by all systems and can be found in figures 1(a) and (b), while the mappings chosen for the gasses are shown in figure 1(c). Groups of atoms that are replaced by a bead throughout the simulation domain are indicated in these figures. All beads are classified with a bead type, which also appears indicated in figure 1 for

ZIF-8, PVDF, $CO_2$, $CH_4$ and $N_2$ individually and beads that replace chemically identical groups of atoms are classified under the same type. As indicated in figure 1(a), two bead types exist for ZIF-8 as beads replace either a Zn cation or an entire ligand. For PVDF (figure 1(b)), four bead types exist. Beads type 1 replace a sequence of two monomers in a chain, beads types 2 and 3 replace the end-chain monomers $CH_3$-$CF_2$- and $CF_3$-$CH_2$-, respectively, and beads type 4 replace a single monomer that may lie between beads of types 1 and 3 in a polymer chain. The latter bead type is required because the PVDF model has a polydispersity such that some of the chains have an odd number of monomers (see figure 1 (d)). Finally, the gasses ($CO_2$, $CH_4$ and $N_2$) are always modelled with one bead replacing each molecule, as shown in figure 1(c). In all cases the masses of the beads are given by the sum of the masses of the atoms they replace.

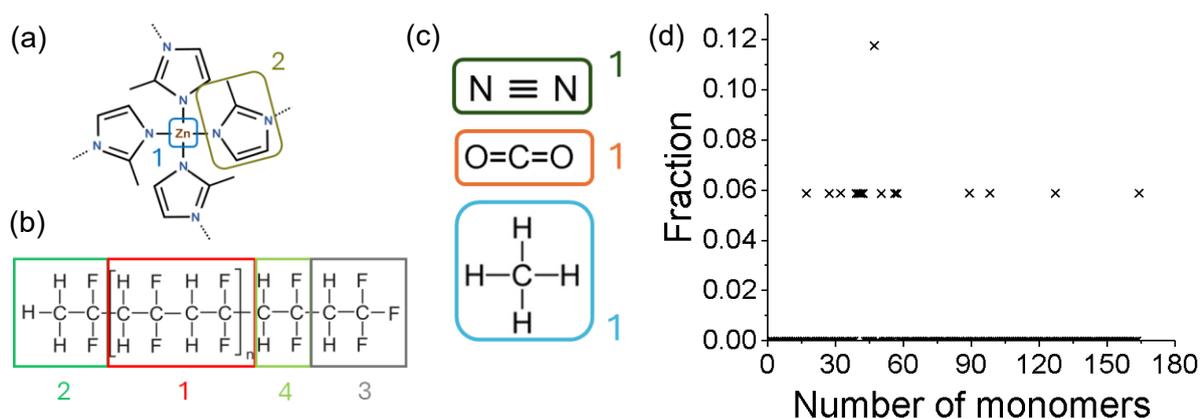

Figure 1. Mappings considered for modelling (a) ZIF-8, (b) PVDF and (c) the three gasses considered in this work, namely, $N_2$, $CO_2$ and $CH_4$. The groups of atoms that are replaced by a bead throughout the system are indicated by superimposing 2D shapes in the all-atom representation of the compounds. The bead type assigned to each bead is also shown for each compound individually. Panel (d) shows the polydispersity considered for the polymer.

The force fields to model the ZIF-8/PVDF/$CO_2$, ZIF-8/PVDF/$CH_4$ and ZIF-8/PVDF/$N_2$ systems in the CG resolution were given by a sum of bonded and pairwise non-bonded potentials and all the potentials that concern only ZIF-8 and/or PVDF beads were taken from the MARTINI/FM force field developed in a previous work to study the ZIF-8/PVDF composite in the absence of gas loading.[32] Non-bonded MARTINI 3 potentials were used for (ZIF-8)-(gas), (PVDF)-(gas) and (gas)-(gas) bead pairs. These follow from the classification of bead flavors made here for each of the three gasses and within the previously mentioned hybrid MARTINI/FM force field for ZIF-8 and PVDF beads (ZIF-8 beads type 1 and 2 were classified as SP5q and SP2aq, respectively, and PVDF beads type 1 had been classified as type RX1 and beads type 2, 3 and 4 as SX1). Fundamentally, the classification within MARTINI relies on how many atoms the beads replace and on how strongly they are expected to interact with water. The $CO_2$, $CH_4$ and $N_2$ beads are classified here as TN3, TC3 and TC1, respectively, following the trend of their reported solubility in water at ambient pressure and temperature conditions (mole fraction solubilities of $6.15 \times 10^{-4}$, $2.55 \times 10^{-5}$ and $1.18 \times 10^{-5}$, respectively).[55] In order to evaluate if the interactions between ZIF-8 and the gas are properly modelled within the previously mentioned force field, CG simulations were made to assess the equilibrium loading and the gas structuring (pair distribution functions for ZIF-8 and gas beads and adsorption sites) for the individual gasses into bulk ZIF-8 at (300 K, 1 atm). The results are presented in the Supporting Information (SI) together with a detailed methodology for these simulations and are quite satisfactory when compared to the experimental references: the experimental number of adsorbed molecules per unit cell lies within the simulation fluctuations (see Ref. 32 and Fig. S1), being very close to the average reference values in the case of $CO_2$, and the adsorption sites are remarkably well captured (see Fig. S2). Note that the ZIF-8/PVDF/$CO_2$ force field had already been proposed and tested in its ability to reproduce the equilibrium $CO_2$ loading in bulk ZIF-8 reported experimentally.[32] The parameters for all the MARTINI potentials involving gas beads deployed in the models for ZIF-8/PVDF/$CO_2$, ZIF-8/PVDF/$CH_4$ and ZIF-8/PVDF/$N_2$ can be found in the SI. Information on potentials involving ZIF-8 and/or PVDF beads can be readily found in the SI of Ref. 32.

II.2) CG simulation setup

Three different ZIF-8 nanoparticles were considered to study the influence of filler morphology and size in the gas adsorption and structuring within the composite: a cube of radius ≈69 Å (CB) and two rhombic dodecahedra of radius ≈62 Å (RD) and ≈28 Å (SRD). An initial configuration for the three ZIF-8/PVDF systems, each comprising only one of these nanoparticles inside a polymer matrix, was borrowed from a previous work.[32] These configurations stem from CG-MD simulations made using the MARTINI/FM force field (see section II.1), the NPT equations of motion with target pressure and temperature set to 1 atm and 300 K and do not contain gas molecules (see SI for more details). It is worth noting that the surface of the CB, RD and SRD nanoparticles feature undercoordinated ligands and Zn ions, which are distributed following a pattern that preserves the expected symmetries of vertices, edges and faces within the geometry of each given nanoparticle. Further useful remarks are that the CB, RD and SRD ZIF-8/PVDF configurations borrowed feature a different composition (%w/w of ZIF-8 follows the trend CB > RD > SRD) from one another, and that the CG model predicts polymer penetration in all cases, featuring ultimately a trend CB > RD > SRD. More details can be found in Ref. 32.

Starting from the initial configurations, two hybrid MC/MD simulations, differing only in the random seed for the MC algorithm, were made for each ZIF-8/PVDF system to study single gas adsorption ($CO_2$, $CH_4$ or $N_2$) using the force fields discussed in the previous section. These hybrid MC/MD simulations featured only gas insertions and deletions as MC moves, which were attempted twice every 10 timesteps, while the dynamics of all beads in the system were carried out using the NVT equations of motion. The target temperature was set to 300 K. A timestep of 1 fs was used for the dynamics aiming to prevent numerical instability issues stemming from the addition and deletion of structural units. The chemical potential associated with the gas reservoir at T = 300 K and P = 1 atm was computed by LAMMPS considering the gas as ideal (i.e., fugacity coefficient = 1).[56] The simulations were carried out until the number of gas molecules oscillated around a mean value, suggesting that the adsorption capacity had been reached. The adsorption capacity for each gas within a given force field and nanoparticle system was determined by averaging the constant mean value of adsorbed gas given by the two independent MC/MD simulations.

The ideal $CO_2$/$CH_4$ and $CO_2$/$N_2$ selectivities, denoted herein as S, were estimated from the values of adsorption capacity attained in the hybrid MC/MD simulations for each nanoparticle system. The calculation follows equation (1), which was also used in previous works.[57,58] In this equation, $A_{gas1}$ and $A_{gas2}$ denote the equilibrium number of molecules of each gas at the given thermodynamic state.

$$S_{gas1/gas2} = \frac{A_{gas1}}{A_{gas2}} \qquad (1)$$

It is important to note that equation (1) does not correspond to the original definition of ideal selectivity, which concerns rather the permeability of the two pure gasses instead of the adsorbed amounts, and therefore is not expected to match it. A qualitative agreement (<1 or >1) is however expected, based on the fact that the diffusion mechanism for the gas inside ZIF-8 should rely on the gas hopping from one adsorption site to another and, therefore, having more adsorption sites for one gas should be correlated with a faster diffusion from a probabilistic point of view.[59] Both $S(CO_2/CH_4)$ and $S(CO_2/N_2)$ values are expected to be favorable towards $CO_2$ adsorption (i.e., > 1) from experimental works.[60-62]

Subsequently, in order to study the equilibrium structuring of the single gas loaded systems, one CG-MD simulation was carried out for the $CO_2$, $CH_4$ and $N_2$ single loaded CB, RD and SRD ZIF-8/PVDF systems, each. These simulations started from a microstate attained in one of the two hybrid MC/MD simulations for the given gas once equilibrium loading had been reached. The NPT equations of motion were used for the dynamics of all the beads, with target temperature and pressure set to 300 K and 1 atm, respectively. The timestep was set to 10 fs. Once fluctuations of potential energy and volume hinted that the system was equilibrated (i.e., fluctuations occur around a constant value), configurations were collected to assess the gas structuring.

ZIF-8, PVDF and gas bead density profiles were computed for the whole simulation domain. Bead density profiles for the polymer and the gas were also built at the surface of the ZIF-8 nanoparticle. The calculations are made precisely in the same way as detailed in a previous work (see Ref. 32), with only an overview given here. The bead density profiles along the simulation domain were built as a function of the distance between the beads and the center of the mass (COM) of the nanoparticle. The polymer and gas bead density profiles at the surface of the ZIF-8 nanoparticle were built considering the position of a bead relative to specific points (referred to as points of assessment throughout the text) set at the vertices, center of the edges and center of the faces featured in the given nanoparticle morphology. More specifically, the profiles were built as a function of the distance (d) between the beads and these points as well as of the smallest angle ($\alpha$) formed between the beads, the point of assessment and the COM of the nanoparticle. Figure 2 illustrates the positioning of the points of assessment for one vertex, edge and face, each, in a configuration for the CB, RD and SRD ZIF-8 nanoparticles. Bead density profiles built for local environments (i.e. vertices, edges and faces) that are equivalent within a given ZIF-8/PVDF system due to the symmetry of the nanoparticle are averaged. Since the rhombic dodecahedron geometry has two groups of vertices that are distinct from one another, vertices of the RD and SRD ZIF-8/PVDF systems are divided into two groups before averaging. Vertices formed by edges that meet in angles of 70.53° and 109.47° are classified into group 1 (g1) and group 2 (g2), respectively. No distinction of bead types was made when building bead density profiles for the ZIF-8 and PVDF phases.

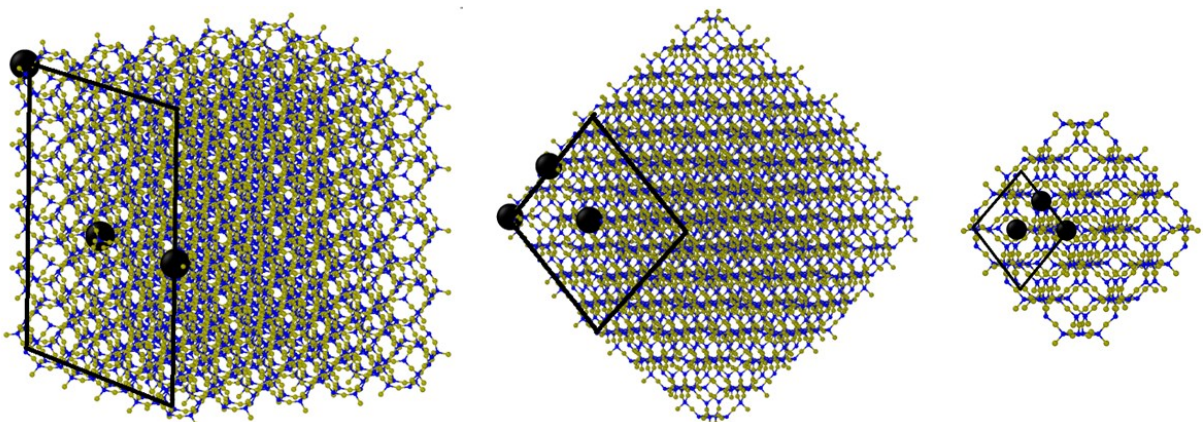

Figure 2. Configuration of the empty CB, RD and SRD ZIF-8 nanoparticles (i.e. without polymer or gas). Beads in blue and dark yellow are beads type 1 and 2, respectively, as defined in figure 1(a). Points of assessment for one vertex, edge and face of each of the nanoparticles are shown in black, and their placing relative to the nanoparticle is defined as briefly discussed in the text and explained more in detail in Ref 32. A face of each of the nanoparticles also appears indicated using black solid lines to aid visualization.

Finally, the total amount of gas beads lying inside the simulation domain was also computed for the single gas loaded CB, RD and SRD ZIF-8/PVDF systems. An in-house code was made to average how many gas beads lie within 48.8 Å of the center of mass of the ZIF-8 CB nanoparticle in the x, y and z directions individually. For the RD and SRD ZIF-8/PVDF systems, the amount of gas beads lying within the simulation domain was estimated by averaging the amount of beads lying within the spheres inscribed and circumscribed to the nanoparticle. In the case of the RD nanoparticle, these spheres have radii of 48 and 62 Å, respectively, while for the SRD nanoparticle, the values are 24 and 28 Å, respectively.

### III) Results

The average equilibrium number of $CO_2$, $CH_4$ and $N_2$ molecules and $CO_2/CH_4$ and $CO_2/N_2$ ideal selectivities obtained for each ZIF-8/PVDF system at ambient conditions are presented in table S2 and figure 3, respectively, together with their error bars. In both, the CB, RD and SRD ZIF-8/PVDF systems are denoted by the type of ZIF-8 nanoparticle in the simulation domain. In figure 3, the selectivities values ≤ 1 are put in bold to highlight systems that fail to meet the higher affinity towards $CO_2$ that is expected for the ZIF-8/PVDF composite based on experimental observations.[60-62] It is worth remembering that the selectivities were calculated considering the number of gas molecules adsorbed in the whole simulation domain. This strictly forbids comparing values in table S2 and figure 3 alone to evaluate the influence of nanoparticle morphology and size on the amount of gas adsorbed, because the composition of the CB, RD, SRD ZIF-8/PVDF systems is different (see section II.1 and Ref. 32).

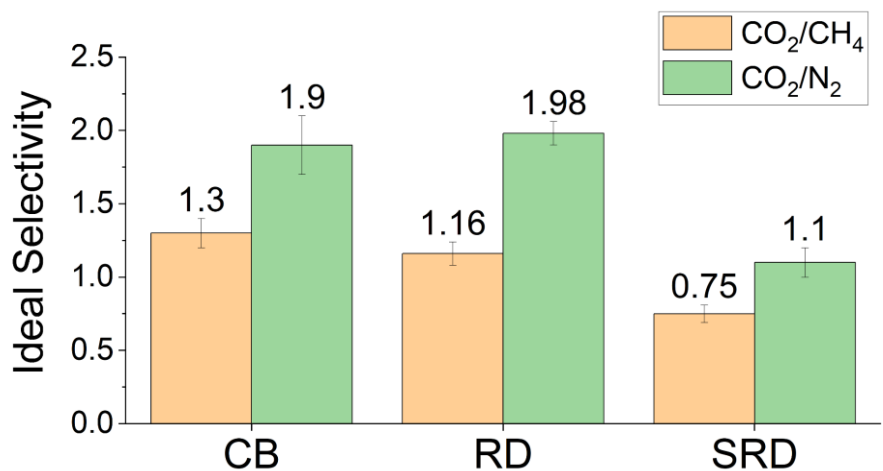

Figure 3. Ideal selectivities, S($CO_2$/$CH_4$) and S($CO_2$/$N_2$), obtained using equation (1) for the ZIF-8/PVDF systems at ambient conditions. The average values are given above the bars to facilitate reading and the uncertainties correspond to the average error.

It is possible to see in figure 3 that the predicted S($CO_2$/$CH_4$) and S($CO_2$/$N_2$) values point towards a modest $CO_2$ favorability in general. This favorability is expected as a consequence of the $CO_2$, $CH_4$ and $N_2$ bead flavors assigned within MARTINI: these bead flavors reflect a polarity trend $CO_2$ > $CH_4$ > $N_2$, which should be passed along to the intensity of the interactions with ZIF-8 beads, as these are classified as highly polar (see Ref. 32 for more information). Thus, provided that the ZIF-8 nanoparticle is indeed the portion of the simulation domain that has a larger amount of gas adsorption sites as expected from experimental observations, the force fields should reproduce the trend of higher favorability towards $CO_2$ compared to $CH_4$ and $N_2$.

The distribution of gas in the nanoparticle domain can be unveiled using bead density profiles built for the simulation domain, which are shown in figure 4 for $CO_2$, $CH_4$ and $N_2$ in the CB, RD and SRD ZIF-8/PVDF systems at (300 K, 1 atm). The x-axis measures the distance between the bead and the COM of the ZIF-8 nanoparticle. The dashed gray lines in the plots mark the largest distances at which ZIF-8 density adopts non-zero values, serving as thresholds to delimit the extension of the nanoparticle in each system. It is possible to see that indeed the force fields predict that gas adsorption in the ZIF-8 phase is significantly larger compared to that in the polymer phase outside the ZIF-8 nanoparticle domain for all three gasses, as previously hypothesized. Notably, figure 4 allows observing larger $CO_2$ bead density values throughout the extension of the SRD nanoparticle domain compared to the RD and CB systems, a difference that is not as significant for $CH_4$ and $N_2$. This would suggest larger values of S($CO_2$/$CH_4$) and S($CO_2$/$N_2$) for the smaller nanoparticle. The low values of selectivity shown in figure 3 could thus be due to a difference in composition. In other words, if several small nanoparticles had been embedded in the nanoparticle domain to even the ZIF-8 loading compared to the big nanoparticles, the selectivities obtained for SRD ZIF-8/PVDF system should be larger than for the CB and RD ZIF-8/PVDF systems. Polymer bead density profiles for the gas loaded systems are not shown here, as they match those obtained for the composite in a previous work in absence of gas loading.[32] Gas bead density profiles at the surface of the nanoparticle are shown in the SI, in which presence of gas can be spotted.

Similarly, only slight modifications are observed in the polymer bead density profiles at the surface level in the gas loaded versus not loaded systems, which are therefore not shown.

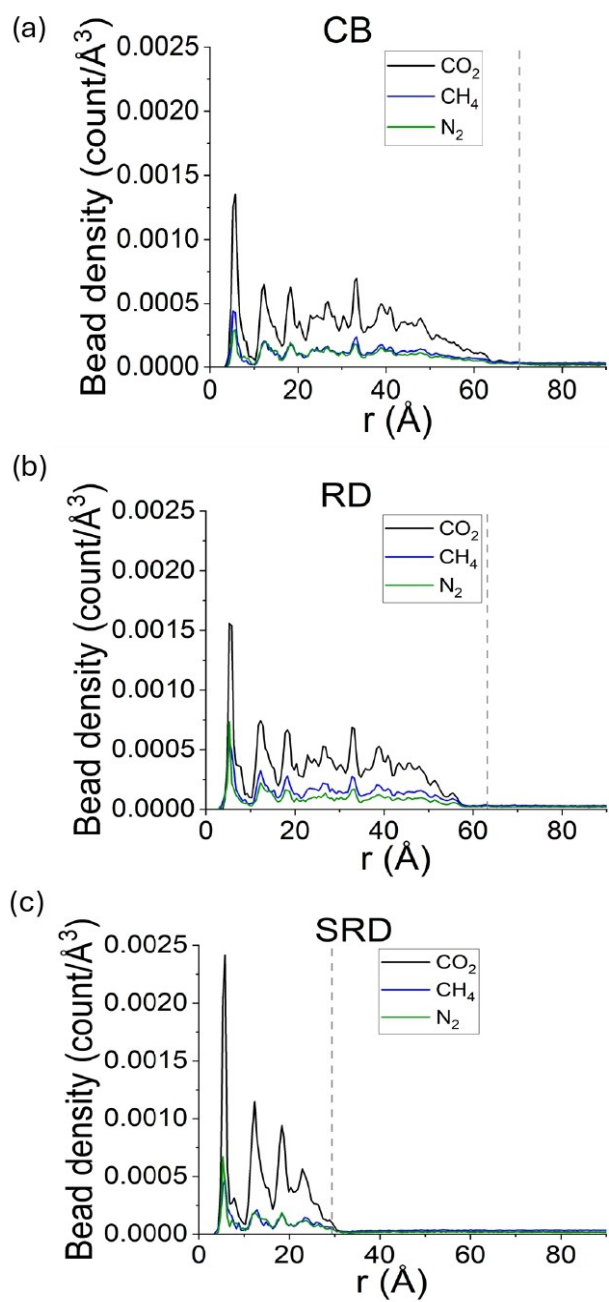

Figure 4. $CO_2$, $CH_4$ and $N_2$ bead density profiles for the single gas loaded (a) CB, (b) RD and (c) SRD ZIF-8/PVDF systems.

Aiming to study the influence of nanoparticle shape and size on gas adsorption, table 1 shows the amount in mg of $CO_2$, $CH_4$ and $N_2$ lying exclusively within the ZIF-8 nanoparticle domain divided by the corresponding mass of ZIF-8 in g. This normalization allows to fairly compare the extent of gas adsorption of the three different nanoparticles investigated, which are formed by a different number of ZIF-8 beads. Additionally, experimental equilibrium amounts of $CO_2$, $CH_4$ and $N_2$ adsorbed in pure ZIF-8 at ambient pressure and temperature conditions are presented in table 1 for reference. For each gas, the experimental reference is the result of averaging the equilibrium loading at 298 K and approximately 1 bar reported in adsorption isotherms of four different works found in the literature.[63-68] This averaging is specifically important in the case of $CH_4$ and $N_2$, since the low equilibrium loading at this (T, P) condition makes it difficult to accurately read the information in the plots even upon digitalization. We note that characterization of the ZIF-8 samples in these experimental works reveals particle diameters ranging from 3-4 $\mu$m[63] to 150-260 nm[65,67,68], except in two cases where no such information was found[64,66]. In addition, in another experimental work, ZIF-8 with average particle size of 105 nm synthesized according to a special protocol has shown a significantly larger equilibrium uptake of $CO_2$, $CH_4$ and $N_2$,[69] which proves that strong deviations from the values shown in table 1 may exist.

|  | $CO_2$ | $CH_4$ | $N_2$ |
|---|---|---|---|
| Experimental reference (pure ZIF-8) | 23 $\pm$ 4 | 5 $\pm$ 2 | 2.3 $\pm$ 0.6 |
| CB | 32.2 $\pm$ 0.8 | 5.0 $\pm$ 0.1 | 5.9 $\pm$ 0.1 |
| RD | 29.4 $\pm$ 0.2 | 4.3 $\pm$ 0.1 | 4.5 $\pm$ 0.1 |
| SRD | 33.5 $\pm$ 0.8 | 2.9 $\pm$ 0.2 | 4.9 $\pm$ 0.4 |

Table 1. Amount of gas adsorbed in the nanoparticle domain (mg of gas per g of ZIF-8) of the CB, RD and SRD ZIF-8/PVDF nanoparticle systems. The uncertainties correspond to the standard errors.

It is possible to see in table 1 that, in general, the force fields predict larger values of gas adsorption at ambient conditions than the experimental reference corresponding to pure ZIF-8. Since the polymer penetration occurring in all ZIF-8/PVDF systems considered is expected to reduce the available space for gas adsorption, the values of adsorption capacity predicted by the models for the composite should be smaller than those for pure ZIF-8 (experimental reference). Thus, there is either an overestimation of the number of adsorption sites and/or of the guest/host interactions strength at the CG level. Indeed, when studying $CO_2$, $CH_4$ and $N_2$ adsorption on bulk ZIF-8 with this same force field, the equilibrium amount of $CO_2$ and $N_2$ adsorbed in bulk ZIF-8 was slightly overestimated compared to the experimental references (see section 1 of the SI). However, disconcertingly, with the exception of $N_2$ adsorption in the RD and SRD ZIF-8/PVDF systems, the equilibrium loading predicted for these two gasses in the composites are significantly larger than the value obtained for bulk ZIF-8 (25 mg/g ZIF-8 and 4.93 mg/g ZIF-8 for $CO_2$ and $N_2$, respectively). A similar behavior is observed for $CH_4$ adsorption, whose loading in bulk ZIF-8 was underestimated compared to the average experimental reference found (3.05 mg $CH_4$/g ZIF-8), in the CB and RD ZIF-8/PVDF systems.

The fact that the equilibrium $CO_2$, $CH_4$ and $N_2$ loadings observed in the simulations for most of the ZIF-8/PVDF composite systems are not only larger than the experimental reference but also larger than what is observed in the simulations for bulk ZIF-8 suggests that a factor other than the overestimation of the strength of (ZIF-8)-(gas) interactions should be contributing to the values obtained in the models and shown in table 1. The larger values observed in the simulations for the composite compared to those for bulk ZIF-8 could be due to the presence of extra adsorption sites at the MOF/polymer interface. Indeed, the gas bead density profiles built at the MOF/polymer interface shown in the SI accuse the presence of gas adsorbed at the MOF surface, which could support the previous statement. Aiming to explore the validity of this hypothesis, the mass of $CO_2$, $CH_4$ and $N_2$ adsorbed in the CB and RD ZIF-8/ PVDF systems was estimated (i) in a sphere inscribed in the ZIF-8 nanoparticle, which has radius of ≈48 Å for both systems, and (ii) in a spherical region of radius 24 Å centered in the center of mass of the nanoparticle, which corresponds to a region very far from the surface (i.e. a bulk-like region). Striving for accuracy in the calculation, only the mass of ZIF-8 lying within these regions was correspondingly used to calculate the loadings. For the same reason, the calculation was also made in a spherical region of radius 24 Å in bulk ZIF-8 to ensure that there are no artifacts due to the geometry of the region considered. The equilibrium loading values (mg gas/g ZIF-8) for all the systems can be found in table 2. Loading values that remain statistically larger or become equivalent to the ones for bulk ZIF-8 within the context of the 24 Å radius sphere are put in bold.

|  | $CO_2$ | $CH_4$ | $N_2$ |
| --- | --- | --- | --- |
| Bulk ZIF-8 simulation | 24 Å: 20 ± 2 | 24 Å: 3.6 ± 0.4 | 24 Å: 4.1 ± 0.6 |
| CB | **24 Å: 30 ± 1**<br>48.8 Å: 32.5 ± 0.4 | 24 Å: 3.1 ± 0.4<br>48.8 Å: 4.2 ± 0.1 | **24 Å: 4.2 ± 0.6**<br>48.8 Å: 4.9 ± 0.2 |
| RD | **24 Å: 33 ± 2**<br>48 Å: 33.0 ± 0.3 | **24 Å: 5.1 ± 0.3**<br>48 Å: 4.5 ± 0.1 | **24 Å: 4.1 ± 0.4**<br>48 Å: 5.1 ± 0.1 |

Table 2. Amount of $CO_2$, $CH_4$ and $N_2$ adsorbed (mg of gas per g of ZIF-8) in spherical regions lying within the ZIF-8 nanoparticle domain for the CB and RD ZIF-8/PVDF systems and the bulk ZIF-8. The radius of the sphere the gas loading concerns is indicated. Spheres of radius 48.8 and 48 Å are those inscribed in the CB and RD nanoparticles, respectively. The uncertainties correspond to the standard errors.

It is possible to see in table 2 that even upon considering gas adsorption in a region very far from the surface (i.e. the 24 Å radius sphere), the CB and RD ZIF-8/PVDF systems contemplate gas loadings equivalent or larger than the one observed for bulk ZIF-8 in the case of most gasses. Since the influence of the surface should be small, if any, in this region, these larger or equivalent loadings observed for the composite mean that the polymer penetration is either not affecting or actually enhancing gas adsorption. The latter scenario, although geometrically possible given that there is plenty of additional space for further adsorption in ZIF-8 at these low gas loading conditions at (300 K, 1 atm), could likely be an unphysical feature of the model. However, our model for the composite at the CG level deploys MARTINI potentials, which are parametrized in a context other than MOF/polymer/gas systems. The fact that the PVDF beads have low polarity (RX1 and SX1) means, within the MARTINI miscibility table,[51] that they are expected to interact well with beads having the low-polarity-bead-flavors assigned to the gasses. This could then explain the high values of gas adsorbed from the point of view of the model: the PVDF lying inside the porosity is promoting further adsorption in the nanoparticle domain and thus leading to the high loadings observed for the composite in table 2. While this means a more careful MARTINI bead flavor assignment may be needed for the hosts, the model can still be considered successful as it prescribes loading values close to expectations. Additionally, it is important to note that the (ZIF-8)-(gas) MARTINI potentials deployed satisfyingly reproduce not only the expected equilibrium loading of the gasses in bulk ZIF-8 compared to the references, but also lead to adsorption sites consistent with what is reported in the literature (see SI), thus supporting the ability of the force field in modelling host-guest interactions.

Given that the loadings in the inscribed spheres in the CB and RD nanoparticles (radius 48.8 and 48 Å, respectively) are expected to suffer from the influence of the surface due to its proximity, comparison with the loadings observed in the spheres of radius 24 Å, which sit far from the surface, offer interesting insights. These values are also given in table 2. In the case of $CO_2$ and $CH_4$, the loading in the inscribed sphere is larger than in the 24 Å radius sphere in the CB ZIF-8/PVDF system. On the other hand, the opposite scenario or no difference is observed for these two gasses in the RD ZIF-8/PVDF system. Given that the two nanoparticles have similar size, this could be associated with differences between the surfaces of the CB and RD nanoparticles. Indeed, the crystallographic planes featured in the surface of these two nanoparticle geometries are not the same: the rhombic dodecahedron and the cubic geometries feature [011] and [100] planes (or their crystallographic equivalents), respectively, at their surfaces. In a computational work previously made by one of the authors, the energies of these two planes in ZIF-8 were calculated and the [011] plane was found to have lower potential energy than the [100] (3.31 *versus* 3.65 J/m$^2$).[31] Thus, higher $CO_2$ or $CH_4$ adsorptions could help stabilize the CB surface. Additionally, the extension (i.e. total area) of the surface in the CB nanoparticle is also larger than in the RD one, given that the two nanoparticles have similar sizes and the former features a higher surface-to-volume ratio than the latter (≈7.0% for the CB *versus* ≈6.2% for the RD). This would contribute further to a larger overall instability and thus higher gas adsorption at the surface level in the CB nanoparticle.

While it is tempting to associate the higher gas adsorption capacity observed for cubic nanoparticles with a better performance for gas-related applications, it is important to remember that selectivity is also a key merit parameter for gas separation. In this matter, it's important to remember that selectivity is calculated based on the number of molecules adsorbed, and not loading in mass units, meaning that the higher loading values for $CO_2$ in tables 1 and 2 could be misleading, as it has a higher molar mass than the other two gasses. Moreover, the fact that the interactions between ZIF-8 and $CO_2$ are stronger compared to the other two gasses does not necessarily mean that more $CO_2$ molecules will be accommodated near the surface of the cubic nanoparticle, since it does not mean less molecules of the other two gasses are required to reach stability. Thus, $CO_2/CH_4$ and $CO_2/N_2$ ideal selectivity values calculated using equation (1) based on the number of gas molecules adsorbed exclusively in the nanoparticle domain (given in table 1) can be used to get a glimpse on how nanoparticle size and morphology affect the performance of the composite in a membrane separation process. These are shown in figure 5. The force field used implies an order of SRD > RD > CB for both S($CO_2/CH_4$) and S($CO_2/N_2$), with the composite having the SRD nanoparticle performing significantly better than the other two in the case of $CO_2/CH_4$. This ultimately means that, for gas separation processes, smaller nanoparticles are expected to perform better for this MOF/polymer pair and that the rhombic dodecahedron geometry is better over the cubic morphology. The former result is aligned with several works that point towards smaller MOF nanoparticles-containing composite membranes yielding better separations.[43-46]

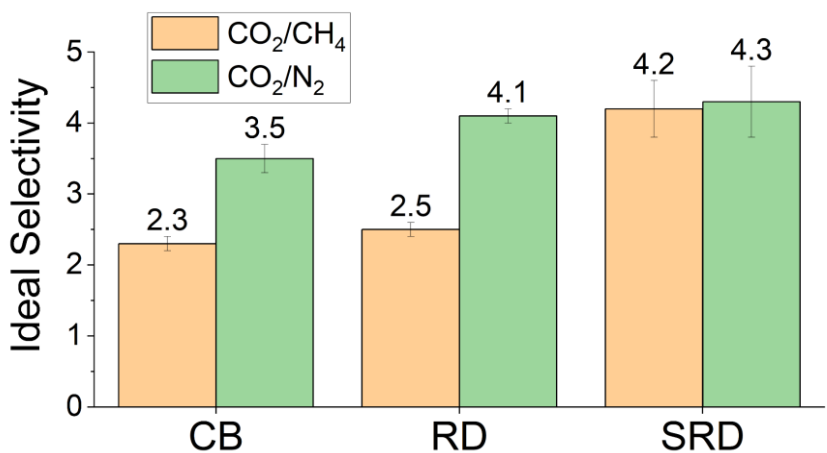

Figure 5. Values of $CO_2/CH_4$ and $CO_2/N_2$ ideal selectivity for the ZIF-8 nanoparticle in each of the systems investigated. Selectivities are estimated counting only the amount of molecules lying within the nanoparticle domain. The average values are given above the bars to facilitate reading and the uncertainties correspond to the average errors computed using the mean values and standard errors of the total amount of adsorbed molecules given in table 1.

## Conclusion

In this work, hybrid MARTINI/FM coarse grained force fields are applied to model $CO_2$, $N_2$ and $CH_4$ adsorption within MOF/polymer composites consisting of a ZIF-8 nanoparticle immersed in a PVDF matrix. In all cases, the intermolecular MOF-polymer interactions are modelled via force matching pairwise potentials. MOF-MOF, polymer-polymer, gas-gas, MOF-gas and polymer-gas interactions are modelled through MARTINI 3 potentials. These force fields perform qualitatively well in depicting the $CO_2/N_2$ and $CO_2/CH_4$ selectivity trends at ambient conditions. At ambient pressure and temperature conditions, higher selectivities are predicted for smaller nanoparticles, making them more appealing for pre- and post- combustion $CO_2$ separations, while the rhombic dodecahedron morphology is preferred over the cubic one. In all cases, gasses are preferentially located within the nanoparticle domain, although there are gas molecules at the bulk polymer phase as well. For individual gasses, it was possible to see that the cubic morphology favors adsorption of gas near the surface of the nanoparticle compared to the bulk, while the rhombic dodecahedron either exhibits no such preference or presents the opposite trend, in particular for $CO_2$ and $CH_4$. This has been ascribed to the fact that the surface of the CB nanoparticle features more unstable crystallographic planes compared to those featured in the RD nanoparticle, which could thermodynamically favor gas adsorption in the former.

This work demonstrates that hybrid MARTINI/FM force fields can give physically sound qualitative information on gas adsorption trends in MOF/polymer composites, allowing to take into account larger systems, so that mesoscopic-level variables, such as morphology and size, can be assessed. This kind of model could further be applied within high-throughput schemes to include these mesoscopic-level variables in the analysis. In addition, this methodology could be extended to model the influence of nanoparticle size and morphology on gas adsorption for other composite materials, thus assisting in designing composites featuring more optimal nanoparticle morphologies and sizes for a given application. We hope this work sparks the interest of the community to model adsorption processes in time- and length scales more amenable to be compared with experimental data via particle-based coarse graining approaches.

## Supporting information

This manuscript is accompanied by a SI which contains specific information (i) on the assessment of the potentials used to model the interactions in the context of bulk ZIF-8/(single gas) systems, (ii) on the configurations for the CB, RD and SRD ZIF-8/PVDF systems that were taken from a previous work to carry out the simulations, (iii) details on the methodology to calculate the gas loading in spherical regions lying within the ZIF-8 nanoparticle domain and bulk ZIF-8 and (iv) figures for gas and PVDF bead density profiles built at the surface of the ZIF-8 nanoparticle for gas loaded ZIF-8/PVDF systems.

Finally, files related to this work are made available at https://github.com/rosemino/MOF_poly_nanoparticle_gas. These files are (i) microstates for the CB, RD and SRD ZIF-8/PVDF systems loaded with $CO_2$, $CH_4$ and $N_2$, each; as well as a microstate for bulk ZIF-8 in the CG resolution considered in our work and (ii) a python code that allows computing the number of gas molecules adsorbed in a spherical region centered in the center of mass of the ZIF-8 nanoparticle.

## Acknowledgements

The authors thank the École Doctoral Sciences Chimiques Balard for funding this work. R.S. thanks the European Research Council for an ERC StG (MAGNIFY project, number 101042514). This work was granted access to the HPC resources of CINES under the allocations A0150911989 and A017091568 made by GENCI.

# Supporting information for: MARTINI-based force fields for predicting gas separation performances of MOF/polymer composites

Cecilia M. S. Alvares,[1] Rocio Semino[2*]

[1] ICGM, Univ. Montpellier, CNRS, ENSCM, Montpellier, France
[2] Sorbonne Université, CNRS, Physico-chimie des Electrolytes et Nanosystèmes Interfaciaux, PHENIX, F-75005 Paris, France
* rocio.semino@sorbonne-universite.fr

Following the same setup as mentioned in the main text, whenever molecular dynamics (MD) simulation details are given throughout the SI, "NVT equations of motion" and "NPT equations of motion" are terminologies used to refer to the Nosé-Hoover equations of motion as originally developed to sample the NVT and NPT ensembles, respectively.[1] The damping constants for the thermostat and (if applicable) for the barostat were of 100x and 1000x the value of the timestep.

1) CG MC/MD simulations for bulk ZIF-8/$CH_4$ and ZIF-8/$N_2$

Aiming to investigate how the potentials involving MOF and/or gas beads perform in reproducing the equilibrium loading at (300 K, 1 atm), one MC/MD simulation was made for ZIF-8/$CH_4$ and ZIF-8/$N_2$ each. ZIF-8/$CO_2$ had already been studied in our previous work.[2] The simulation starts from a bulk CG ZIF-8 configuration contemplating a density that is the equilibrium one in the absence of gas loading at (300 K, 1 atm) according to the force field. The bead dynamics is carried out using the NVT equations of motion and a total of 2 gas insertions or deletions are attempted as Monte Carlo (MC) moves every 10 timesteps. The timestep was set to 1 fs in order to prevent numerical instabilities due to the insertion/deletion of gas molecules. The fugacity coefficient of both gasses, used by LAMMPS to compute their chemical potential within the scope of MC at the grand canonical ensemble,[3] was considered to be 1. The simulation spans a total length of 5 ns, observed to be sufficient to reach the equilibrium loading at the given (T,P) condition.

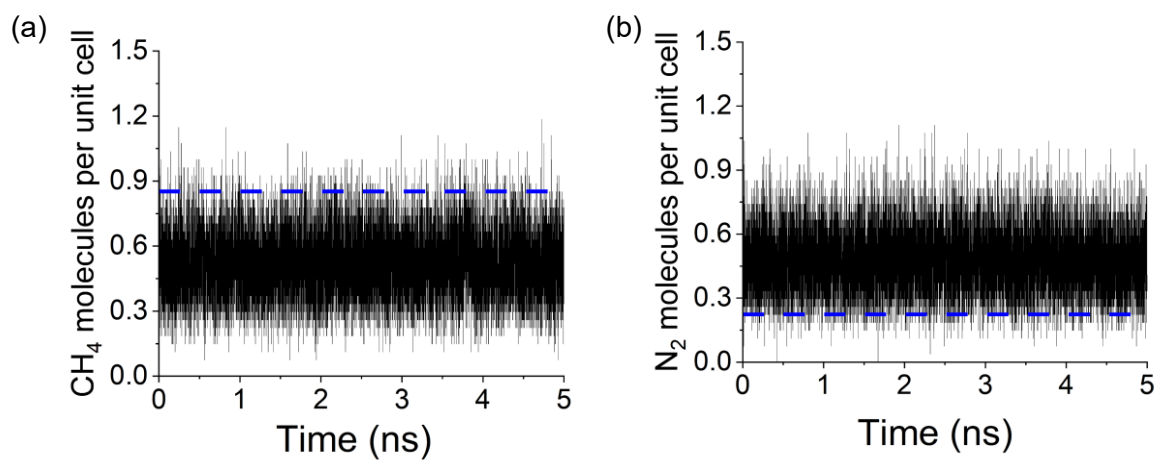

Figure S1. Number of (a) $CH_4$ and (b) $N_2$ molecules per unit cell found throughout the hybrid MC/MD simulation made for each gas. Dashed horizontal lines mark the experimental

references observed for pure ZIF-8 in this condition. The reader is referred to Ref. 2 for a similar plot for $CO_2$.

Figure S1 shows the number of gas molecules adsorbed per unit cell as a function of the simulation time obtained from the CG simulations. A horizontal dashed blue line indicates the experimental reference for pure ZIF-8, taken from adsorption isotherms found in the literature, at (300 K, 1 atm) for $CH_4$ and $N_2$.[4-9] These values are the same as presented in table 1 of the main text, except that here they are converted to molecules per unit cell. It is possible to see that the models predict an equilibrium loading at (300 K, 1 atm) that has some offset compared to the experimental references reported for both gasses, but with fluctuations that encapsulate the reference value. This was also observed in our previous work for $CO_2$,[2] except that the offset was slightly smaller: the average $CO_2$ loading accused in the simulation is 1.52 $CO_2$ molecules per ZIF-8 unit cell (or 25 mg $CO_2$/g ZIF-8) *versus* 1.43 $CO_2$ molecules per ZIF-8 unit cell (or 23 mg $CO_2$/g ZIF-8) (see Ref. 2 for more information). In the context of $CH_4$ and $N_2$, it is important to point out the large scattering of experimentally reported equilibrium loading at ambient (P,T) conditions read in the digitalized adsorption isotherm plots: values ranging from 0.47 to 1.2 $CH_4$ molecules per ZIF-8 unit cell (or 2.78 to 7.07 mg $CH_4$/g ZIF-8) and 0.16 to 0.28 $N_2$ molecules per ZIF-8 unit cell (or 1.69 to 2.88 mg $N_2$/g ZIF-8) have been found.[4-9] Taking into account this large variation, the average $CH_4$ loading in bulk ZIF-8 observed in the simulation (0.52 molecules per unit cell or 3.05 mg $CH_4$/g ZIF-8) can be considered sufficiently good, as it lies within the experimental values near to the lower limit. The same cannot be said for $N_2$, though, whose average value accused in the simulation is 0.48 molecules per unit cell or 4.93 mg/ g ZIF-8. Since there are no less polar beads to assign to this gas within the MARTINI 3 classification, it would be necessary to modify the ZIF-8 beads flavors for reproducing its adsorption. Yet, depending on the level of accuracy desired for reproducing equilibrium loading, even the results observed here for $N_2$ may be considered sufficiently good.

The stable adsorption sites of each of the gasses as well as radial distribution functions (RDFs) between ZIF-8 and gas beads were also assessed. In this matter, for each of the three gasses, a MD simulation for bulk ZIF-8/gas was made using the NPT equations of motion where target pressure and temperature were set to 1 atm and 300 K. These simulations started from the last configuration attained in the hybrid MC/MD simulation to study gas adsorption and the timestep of the hybrid simulation was kept. Note that in the case of $CO_2$, the configuration comes from the hybrid MC/MD simulation made in our previous work.[2] These simulations contemplate 2 ns long equilibration and production, each. A total of 2000 configurations were collected every 1 ps during the production to build the RDFs, while the assessment of the adsorption sites for the gasses within bulk ZIF-8 was made using the final configuration attained by the end of the production. Furthermore, configurations were collected during the production for the assessment discussed in section 4 below.

Figure S2 shows the RDFs built for ZIF-8 and gas beads as well as the final configuration attained in the MD simulations made for bulk ZIF-8/gas for each of the three gasses, where relevant molecules are highlighted and their surroundings are shown in detail for discussion. As a reference, in the case of $CO_2$, a pioneering work in which DFT calculations were deployed to estimate the most important adsorption sites in ZIF-8 revealed that the most relevant ones were (i) near the methyl group of the ligand and (ii) near the 6-MR window.[10] For $CH_4$, theoretical and experimental works conducted at low temperatures (125 K and below) have reported that the adsorption sites at low loadings were located (i) in the center of the 6 MR windows and (ii) on the top of the C=C bond of the ligand.[11,12] In the latter case, based on distance measurements made available by the authors and on an estimate of the C=C bond length,[13] the distance between the center of the C=C bond and the closest hydrogen of the methane molecule is between 3.4 and 4.2 Å. Finally, a computational work made to study $N_2$ adsorption on ZIF-8 at 77 K revealed that the adsorption sites for $N_2$ are the same evidenced experimentally in the case of $CH_4$ at loadings up to 36 $N_2$ molecules per unit cell.[14,12]

It is evident that some of the adsorption sites found in the literature for the gasses in ZIF-8 cannot be reproduced properly at the CG level since the whole ligand is coarsened into a bead, which causes the loss of the specific position of the methyl group and C=C bond. Yet, for all three gasses it is possible to spot molecules in sites consistent with what is expected within the CG resolution considered here. In figures S2(a), S2(c) and S2(e), two molecules, each found in a spot consistent with one of the two adsorption sites expected for a given gas, are highlighted.

Specifically in the case of $CH_4$, the highlighted molecule that is found next to a 4-MR window is at a distance of $\approx$ 4 Å, 4.4 Å, 5.2 Å and 5.3 Å from each of the ligands forming the 4MR window, and > 6 Å from the other two ligands evidenced in the surroundings in figure S2(c). Given that the ligand and the $CH_4$ molecules are coarsened in a single bead each in our model, these values are aligned with what is expected to be observed in the coarsened resolution for the all-atom site near the C=C bond of the ligand. This is because larger values than the references mentioned in the previous paragraph (3.4 to 4.2 Å) for this site are expected to occur if the distance between the center of masses of the ligand and of the $CH_4$ molecule at this adsorption site is considered instead: in this case, it is necessary to account for the C-H bond length in methane and for the distance between the center of mass of the ligand and the center of the C=C bond. The final possible values are also expected to be impacted by the many different rotations of the ligand.

Molecules that sit in positions consistent with expected reference adsorption sites can also be found for $CO_2$ and $N_2$. These are highlighted in figure S2(a) and S2(e) for $CO_2$ and $N_2$, respectively. The $CO_2$ molecule lying near a 4-MR window sits at $\approx$ 3.8, 4.2, 5.5 and 5.7 Å from the ligand beads forming this window, while the $N_2$ molecule in the same scenario is at $\approx$ 3.7, 4.9, 5.4 and 6.7 Å from the ligands forming the 4-MR window. It is interesting to note that the two smallest (gas)-(ligand) distances found near the 4-MR environment are on average smaller in the case of $CO_2$ than in the case of $CH_4$ and $N_2$. While this could be a coincidence due to the dynamics, the shortest distance between $CO_2$ molecules and ligands could be associated with the fact that the center of mass of the ligand sits closer to the methyl group, which is a $CO_2$ adsorption site, than to the C=C bond, which is the preferred adsorption site of both $CH_4$ and $N_2$ molecules. The fact that the mode of the first peak in the RDF Ligand-$CO_2$, shown in figure 2(b), is at a lower distance ($\approx$ 4.0 Å) than that of the Ligand-$CH_4$ ($\approx$ 4.1 Å) and Ligand-$N_2$ ($\approx$ 4.2 Å) helps support this hypothesis and thus shows the good quality of the CG force field used to model the interactions in rightfully depicting the gas structuring despite the high level of coarsening proposed in this work.

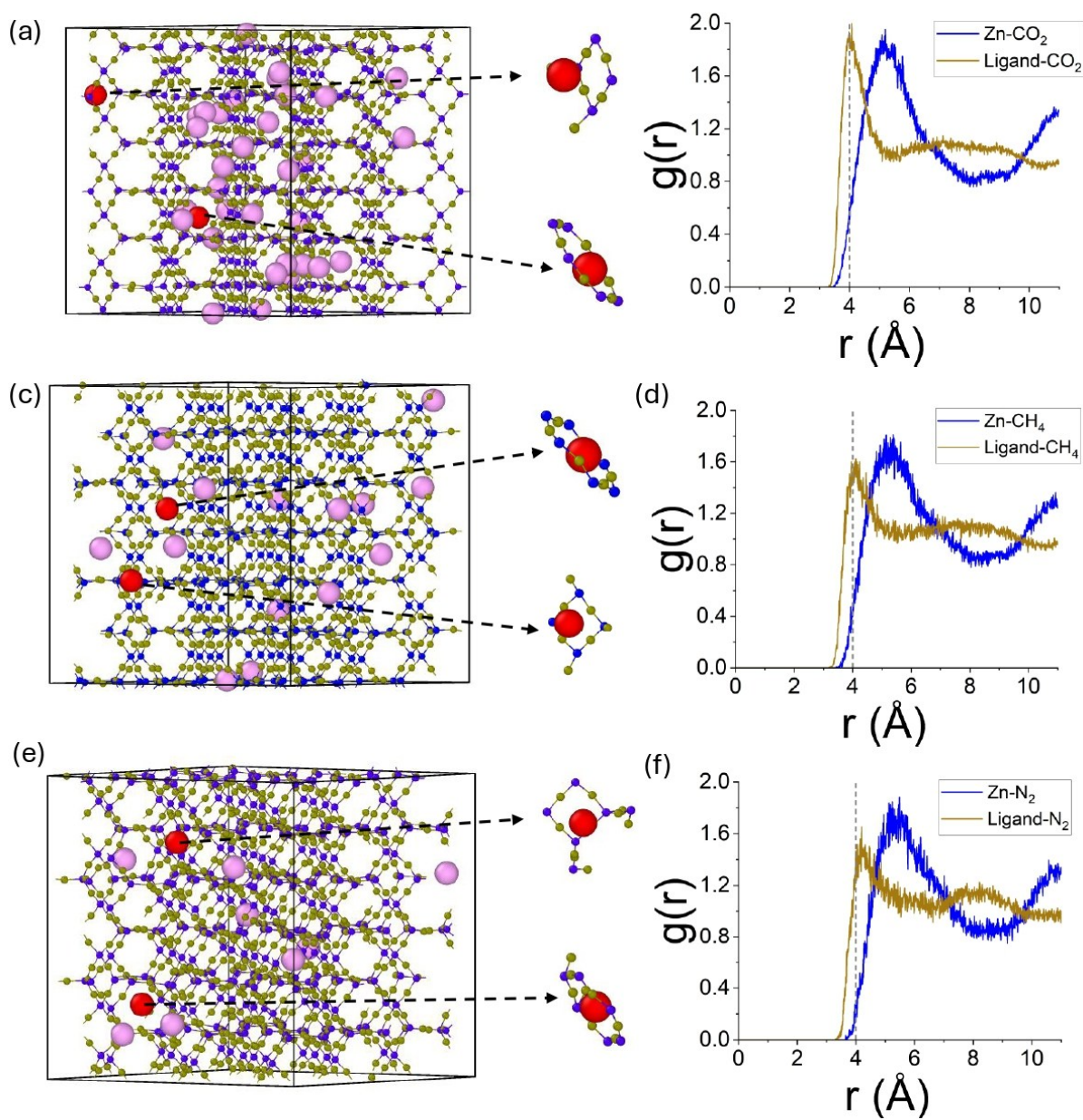

Figure S2. Snapshots of the configurations attained by the end of the production run in the simulation for bulk ZIF-8/gas where ZIF-8 beads type 1 (Zn beads) and 2 (ligand beads) are shown in blue and dark yellow, respectively, and the gas beads are shown in light pink (see mapping in figure 1 of the manuscript) for (a) $CO_2$, (c) $CH_4$ and (e) $N_2$. Two gas beads that lie in positions consistent with the expected adsorption sites for each of the gases based on previous works are highlighted in red and have their environment (or surroundings) shown in the right hand side. Radial distribution functions built for (ZIF-8)-(gas) beads in the simulation for (b) $CO_2$, (d) $CH_4$ and (f) $N_2$. All three plots have the same x and y axis scale and the color of the lines match the color used for the given ZIF-8 beads in the snapshots. Additionally, dashed gray lines with equation r = 4 Å appear in the RDF plots to assist with the visualization of peak mode position.

2) MARTINI 3 potential parameters

Table S1 compiles the parameters of all non-bonded MARTINI potentials involving gas beads used in the ZIF-8/PVDF/gas models assembled to study the composite under single gas loading. These stem from the bead flavor classification made for ZIF-8, PVDF, $CO_2$, $CH_4$ and $N_2$ beads within MARTINI 3.[15] All other potentials involved in these models come from the hybrid MARTINI/FM force field developed to study ZIF-8/PVDF in the absence of gas loading and can be found at Ref. 2.

|  | $CO_2$ | $CH_4$ | $N_2$ |
|---|---|---|---|
| ZIF-8 beads type 1 | $\varepsilon = 0.5043$ <br> $\sigma = 3.65$ | $\varepsilon = 0.34655$ <br> $\sigma = 3.78$ | $\varepsilon = 0.308$ <br> $\sigma = 3.94$ |
| ZIF-8 beads type 2 | $\varepsilon = 0.5593$ <br> $\sigma = 3.65$ | $\varepsilon = 0.38241$ <br> $\sigma = 3.66$ | $\varepsilon = 0.3465$ <br> $\sigma = 3.78$ |
| PVDF beads type 1 | $\varepsilon = 0.5043$ <br> $\sigma = 3.95$ | $\varepsilon = 0.5521$ <br> $\sigma = 3.95$ | $\varepsilon = 0.5043$ <br> $\sigma = 3.95$ |
| PVDF beads type 2, 3 and 4 | $\varepsilon = 0.4183$ <br> $\sigma = 3.65$ | $\varepsilon = 0.4565$ <br> $\sigma = 3.65$ | $\varepsilon = 0.4183$ <br> $\sigma = 3.65$ |
| $CO_2$/$CH_4$/$N_2$ | $\varepsilon = 0.423$ <br> $\sigma = 3.4$ | $\varepsilon = 0.3609$ <br> $\sigma = 3.4$ | $\varepsilon = 0.3609$ <br> $\sigma = 3.4$ |

Table S1. Parameters of the MARTINI 3 non-bonded potentials deployed to study ZIF-8/PVDF single gas loaded with $CO_2$, $CH_4$ and $N_2$. Each column carries information about the system containing a specific gas ($CO_2$, $CH_4$ or $N_2$) and the lines specify the parameters of the non-bonded potentials of the given gas with ZIF-8 and PVDF bead types as well as with the gas itself (last line of the table, in which the non-bonded potential parameters for $CO_2$-$CO_2$, $CH_4$-$CH_4$, $N_2$-$N_2$ are duly given).

3) Further details on the configurations borrowed from previous works for the CB, RD, SRD ZIF-8/PVDF systems.

The initial configurations for the CB, RD and SRD ZIF-8/PVDF systems used in this work correspond to the ones attained after 78 M, 78 M and 30 M timesteps, respectively, in CG-MD simulations made in a previous work using the NPT equations of motion with target pressure and temperature set to 1 atm and 300 K.[2]

4) Assessment of gas loading in spherical regions

Following the discussion in the Results section of the main manuscript, $CO_2$, $CH_4$ and $N_2$ loadings (mg gas per g of ZIF-8) were assessed in spherical regions lying within the nanoparticle domain for the CB and RD ZIF-8/PVDF systems in order to better understand the origins of the loading values estimated within the ZIF-8 nanoparticles (shown in table 1 of the main manuscript). A total of 500 configurations collected after the system had equilibrated were considered in the calculation, which was done using an in-house python script. The script first (i) ensures the MOF nanoparticle is unwrapped, (ii) computes the coordinates of its center of mass and (iii) translates, as necessary, all polymer and gas beads across periodic boundaries so that they sit in the domain closest to the center of mass of the MOF. Then, the script computes the average number of ZIF-8 and gas beads lying inside a sphere centered in the center of mass of the nanoparticle for each group of 20 configurations. The ZIF-8 beads types 1 and 2 were distinguished in the calculation and two spherical regions were considered: one with radius of 24 Å and another that corresponds to the sphere inscribed in the corresponding nanoparticle. The latter has radius 48.8 and 48 Å for the CB and RD ZIF-8/PVDF systems, respectively. Since a total of 500 configurations split into groups of 20 were considered, 25 average values of ZIF-8 and gas beads lying inside the spherical regions were computed. These were treated as samples in order to estimate the standard error when calculating the gas loading in the spherical regions. Specifically, the calculation for the spherical region of radius 24 Å was also done for the bulk ZIF-8 simulations using 500 configurations collected during the MD simulation discussed in section 1. This was done in order to have a reference for this system considering a region with the same geometry, which may differ slightly in number of ZIF-8 beads and thus, consequently, in gas loading compared to the cubic simulation domain, used to estimate the reference gas loading values in section 1. The in-house python script used in the calculation in this case considers the center of the simulation domain as the center of the 24 Å radius sphere and the same strategy of splitting the 500 configurations into 25 groups for the sake of statistics was deployed.

5) Results: loading in the composite and bead density profiles at the surface level

Table S2 shows the average number of $CO_2$, $CH_4$ and $N_2$ molecules found adsorbed in the whole simulation domain in the CB, RD, SRD ZIF-8/PVDF systems at equilibrium at (300 K and 1 atm).

|  | $CO_2$ | $CH_4$ | $N_2$ |
| --- | --- | --- | --- |
| CB | 445 ± 37 | 353 ± 6 | 235 ± 1 |
| RD | 415 ± 8 | 357 ± 18 | 210 ± 4 |
| SRD | 189 ± 11 | 251 ± 6 | 165 ± 6 |

Table S2. Equilibrium number of $CO_2$, $CH_4$ or $N_2$ molecules adsorbed in the single gas loaded ZIF-8/PVDF nanoparticle systems at ambient conditions. The uncertainties correspond to the standard errors.

Figures S3, S5 and S7 show the $CO_2$ bead density profiles for the CB, RD and SRD ZIF-8/PVDF/$CO_2$ systems, respectively, while figures S4, S6 and S8 show the same for the ZIF-8/PVDF/$CH_4$ systems. Figures are not shown for the ZIF-8/PVDF/$N_2$ system because the profiles greatly resemble those for ZIF-8/PVDF/$CH_4$. All values of density in the contour plots are linear density values, and the bead density profiles for the SRD ZIF-8/PVDF system span a smaller distance range than the CB and RD ZIF-8/PVDF systems (5 Å versus 12 Å). The reasons underlying each of these two features can be found in the complete description of the methodology for the calculation, made in Ref. 2.

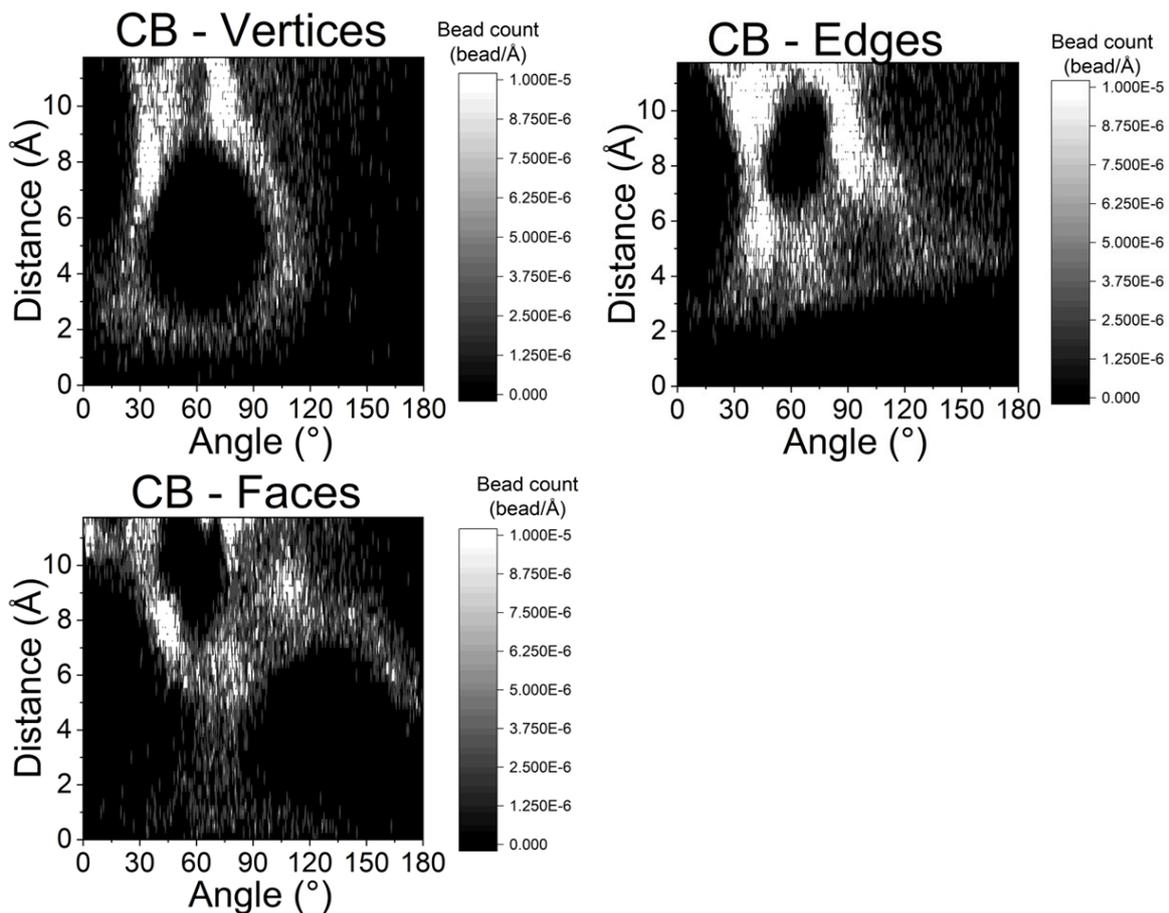

Figure S3. $CO_2$ bead density profiles for the ZIF-8/PVDF/$CO_2$ CB system. In the contour plot, values in white are $\geq 1\times10^{-5}$.

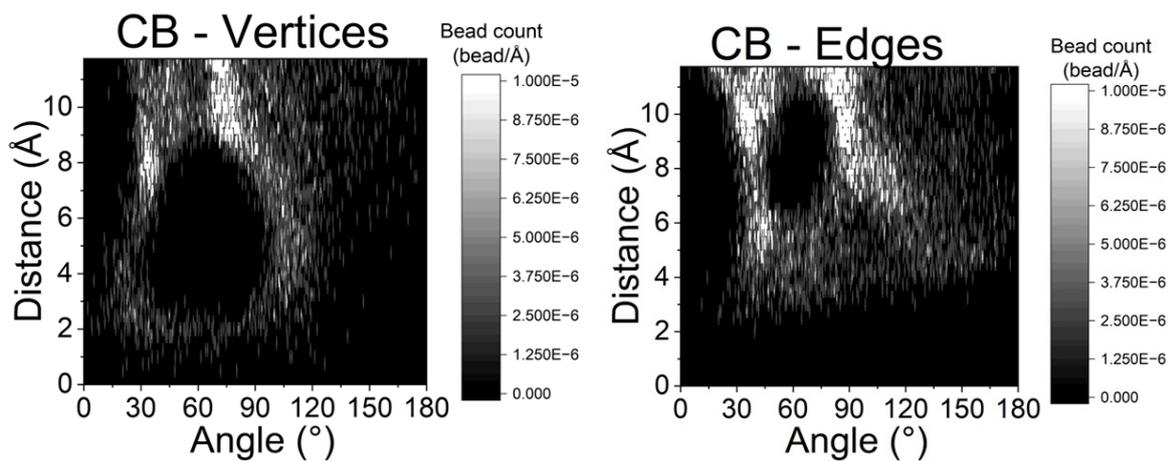

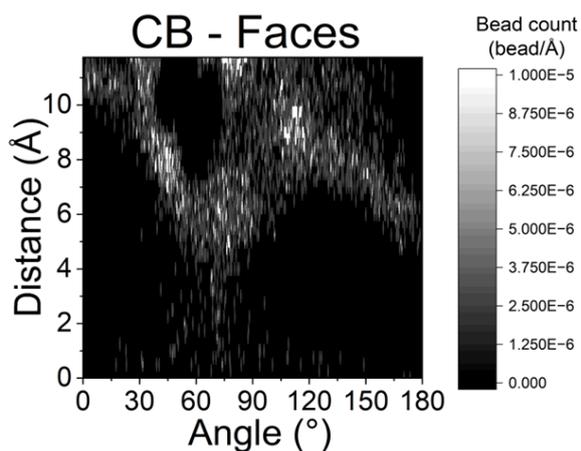

Figure S4. CH$_4$ bead density profiles for the ZIF-8/PVDF/CH$_4$ CB system. In the contour plot, values in white are ≥ 1x10$^{-5}$.

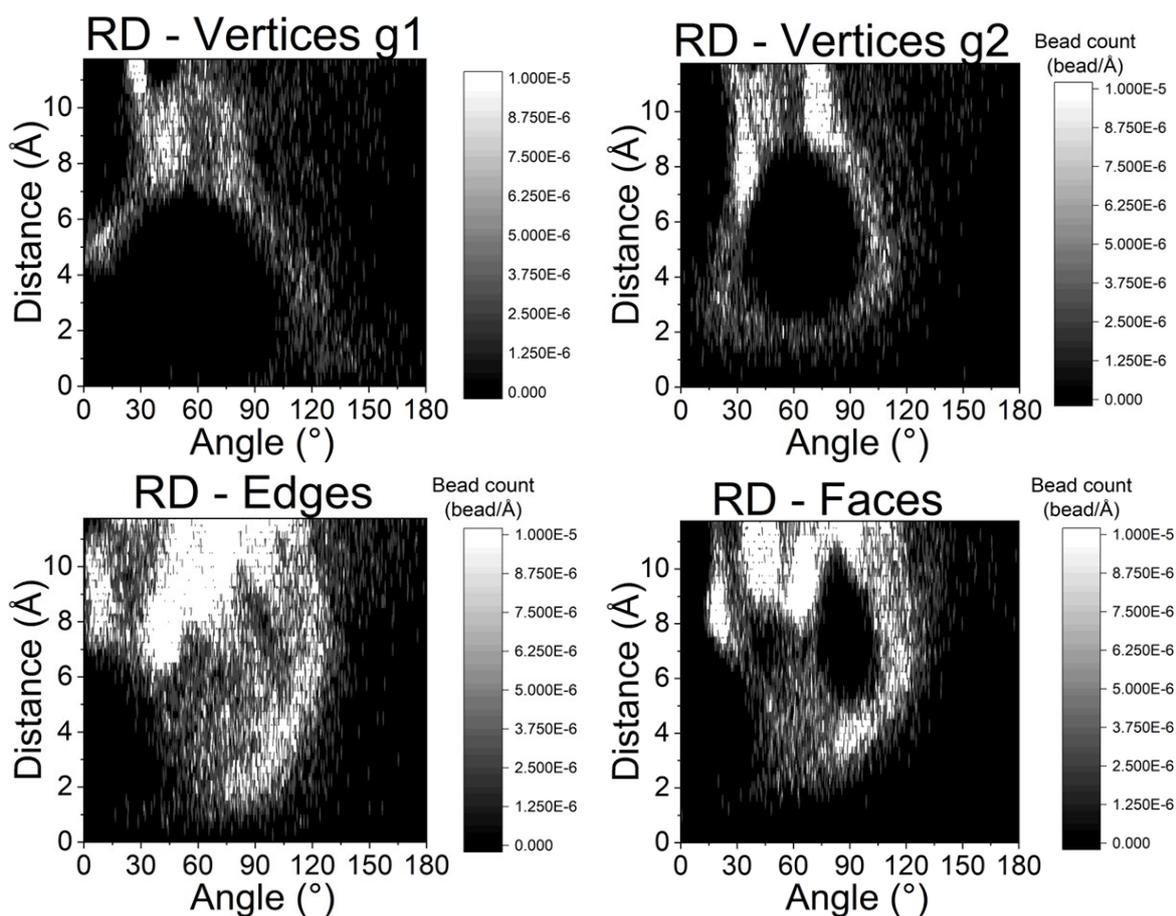

Figure S5. CO$_2$ bead density profiles for the ZIF-8/PVDF/CO$_2$ RD system. In the contour plot, values in white are ≥ 1x10$^{-5}$.

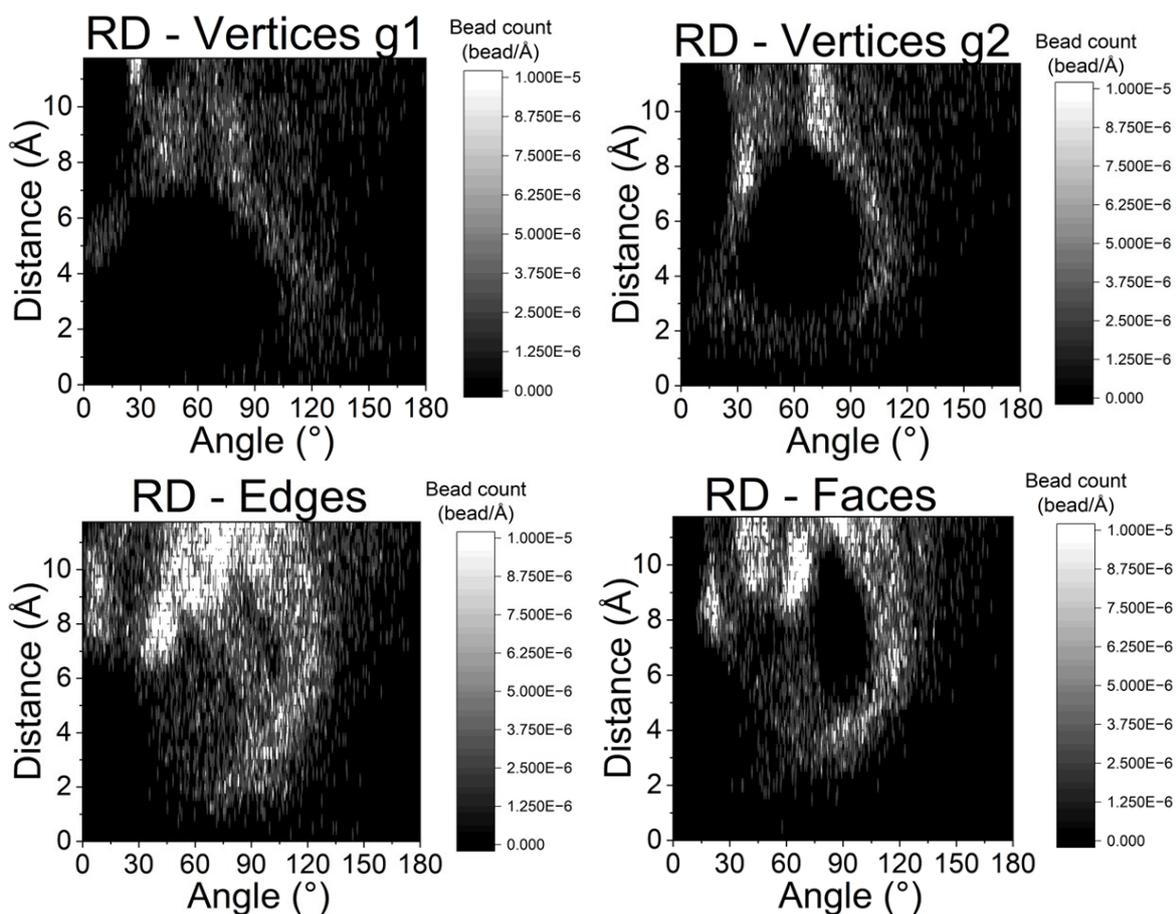

Figure S6. CH$_4$ bead density profiles for the ZIF-8/PVDF/CH$_4$ RD system. In the contour plot, values in white are ≥ 1x10$^{-5}$.

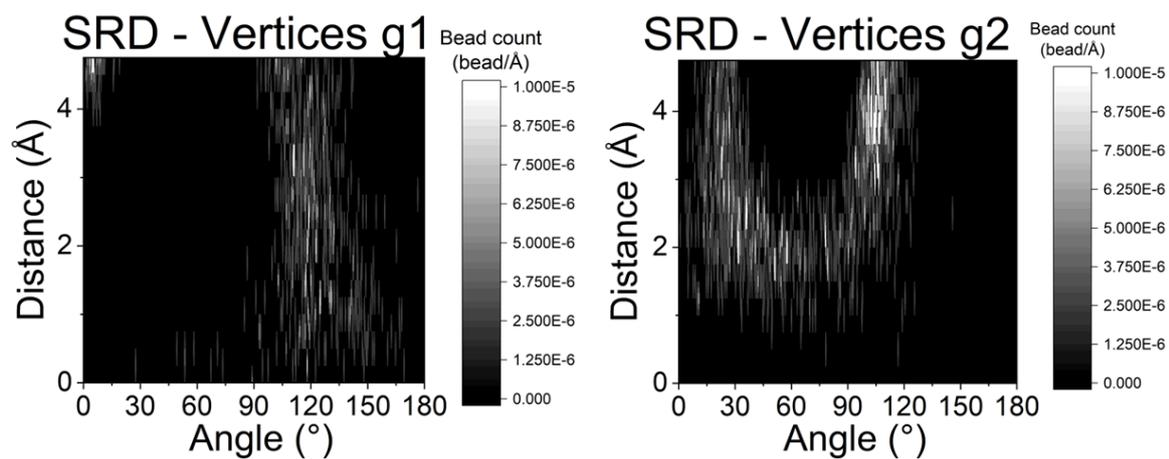

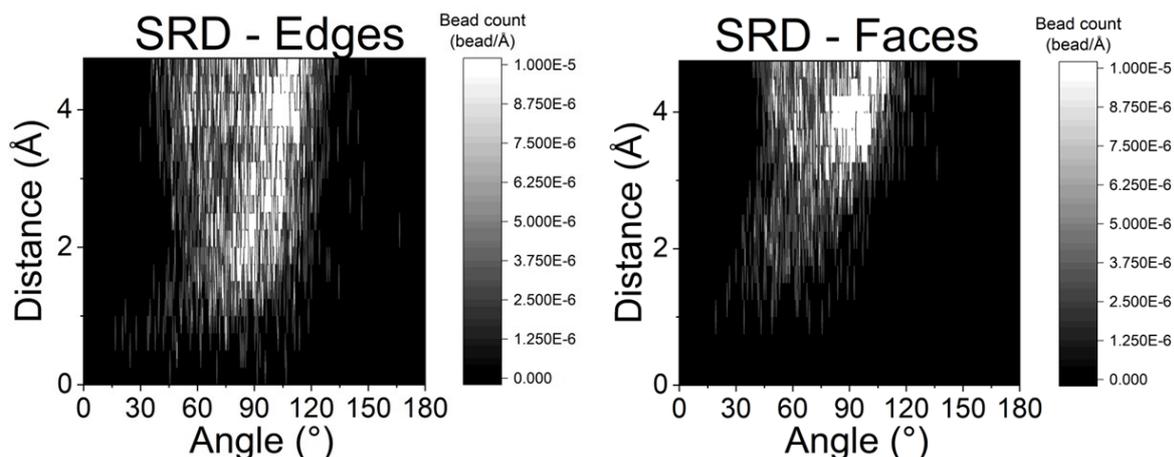

Figure S7. $CH_4$ bead density profiles for the ZIF-8/PVDF/$CO_2$ SRD system. In the contour plot, values in white are $\geq 1 \times 10^{-5}$.

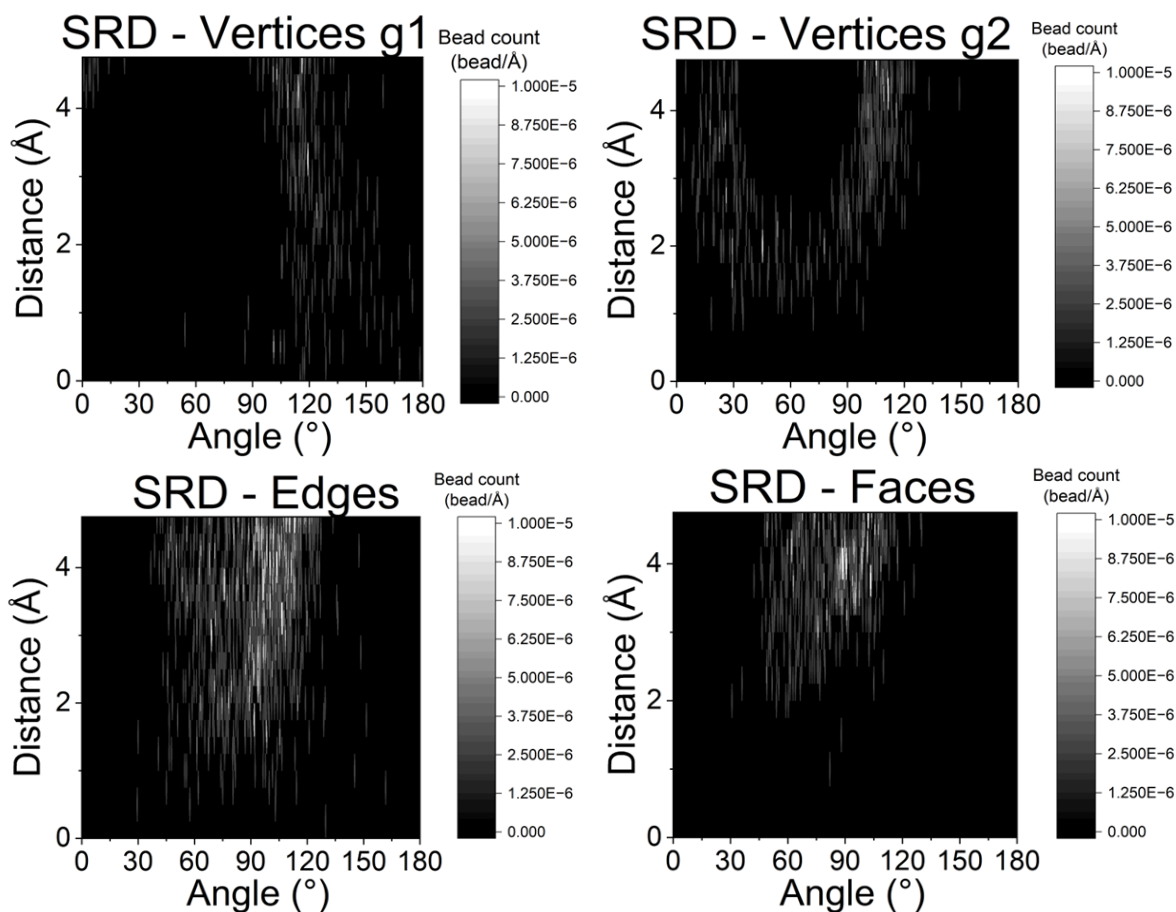

Figure S8. $CH_4$ bead density profiles for the ZIF-8/PVDF/$CH_4$ SRD system. In the contour plot, values in white are $\geq 1 \times 10^{-5}$.